\documentclass[amsmath,amssymb,twocolumn,showpacs]{revtex4}
\usepackage{amsfonts}
\usepackage{mathrsfs}
\usepackage{graphicx}  
\usepackage{cases}

\begin{document}

\title{A spin-boson theory for charge photogeneration in organic molecules: Role of quantum coherence}

\author{Yao Yao}
\affiliation{State Key Laboratory of Surface Physics and Department
of Physics, Fudan University, Shanghai 200433, China}

\date{\today}

\begin{abstract}
The charge photogeneration process in organic molecules is
investigated by a quantum heat engine model, in which two molecules
are modeled by a two-spin system sandwiched between two bosonic
baths at their own temperatures. The two baths represent the photon
emission source and the phonon environment, respectively. We utilize
the time-dependent density matrix renormalization group algorithm to
investigate the ultrafast quantum thermodynamical processes of the
model. We find that the transient energy flow through the two spins
behaves a two-stage effect: The first stage shows a coherent
dynamics which represents the ultrafast delocalization and
dissociation of the charge-transfer state, and in the second stage a
steady current is establish. The photo-to-charge conversion is highly efficient with the maximum efficiency being $93\%$
with optimized model parameters. The survival entanglement between
the two spins is found to be mostly responsible for the hyper
efficiency.
\end{abstract}


\maketitle

\section{Introduction}

In nature, the charge photogeneration takes place in the
light-harvesting systems such as green sulfur bacteria, in which the
photon-to-charge conversion efficiency is remarkably high as in the
extreme conditions the efficiency could be approximately $100\%$
\cite{LHC1,LHC2,LHC3}. It is recognized that a long-lived quantum
coherence gives rise to the hyper efficiency. The time-resolved
two-dimensional optical spectroscopy experiment in
Fenna-Matthews-Olson antenna complex revealed that the coherent time
is incredibly longer than $400{\rm ps}$ \cite{LHC4}. On the
contrary, however, the artificial photocells based upon the organic
molecules merely produce the efficiency around $10\%$ \cite{bredas}.
The deleterious factors in the molecular materials such as the
disorders and the traps break the quantum coherence, leading to
significant reduction of the charge carrier's mobility and the
conversion efficiency \cite{book}. In this context, people devoted
ever-growing efforts in the last two decades to minimize the
deleterious process in molecular photocells in order to get higher
conversion efficiency. The essential contributions rely on the
polymer-based solar cells in terms of the bulk heterojunction (BHJ)
structure along with high-conductivity polymers \cite{BHJ}. A
long-term interest of the community is then to uncover the intrinsic
working mechanism of BHJ photocells.

The process of photogeneration in the BHJ photocells consists of
three steps \cite{review}: (i) A donor molecule (polymer) staying at
the ground state is excited by the photon energy to form an exciton,
which moves and quickly reaches the donor-acceptor interface due to
the structure of BHJ. (ii) Through an ultrafast transition process,
the electron in the exciton transfers into the neighborhood acceptor
molecule (fullerene or its derivative) and forms a charge-transfer
(CT) state along with the hole staying in the donor molecule. (iii)
The electron further moves away from the hole and becomes a free
charge carrier, producing the useful work of the photocell. The CT
state which only exists in the organic photocells is essential for
the charge photogeneration. It is the origin of free charges and
useful work, but it may at any time geminately recombine and lose
the excitation energy to the environment making the useful work and
efficiency decrease. As the dielectric constant of organic materials
is rather small (about $3-5$), the mean distance between the
electron and hole in the CT state is thought to be short, which
induces a relatively large Coulomb attractive energy. Intuitively,
it then seems more likely for the CT state to recombine rather than
to dissociate to free charges \cite{dissModel}. Hence, the possible
ingredient of the driving force for the dissociation of CT state is
hotly discussed. The possibilities include the electric field
\cite{Gregg1}, the energy disorder and entropy driving force
\cite{Gregg2}, the excess energy that from exciton to CT state
\cite{Troisi1}, and the charge or exciton delocalization
\cite{delo}.

Recent experimental researches focused on the delocalization of
wavefunctions in the polymers and the role of quantum coherence
\cite{Friend1,Friend2,Friend3,Guo,ScienceNew}. A pump-push technique
was utilized, in which the push pulse is to make the CT state
repopulate and enhance the population of the relevant hot CT state
\cite{Friend1}. Due to the delocalization and the energetics, the
hot CT state is easy to be dissociated. This statement was
demonstrated by the observations that the decay of the transient
photocurrent could be greatly slowed down. Further experiment based
on the transient absorption spectroscopy investigated the role of
spin in the recombination \cite{Friend2}. It was motivated by the
realization that, if the electron and hole in the CT state are bound
with each other very tightly the geminate recombination will be
dominant, otherwise the bimolecular recombination matters. The spin
participates only in the latter case which is essentially observed
in the experiment, implying the significant role of the bimolecular
recombination and thus the CT state delocalization. A recent
experiment further contributed to the role of electric field that is
induced by the separation of the electron and hole \cite{Friend3}.
Accordingly, the boundary of the coherent and localized phase was
established which is critically important for the study of quantum
coherence in organic photocells. In addition, another group also
made use of the transient photoinduced absorption measurement to
study the BHJ photocells made by various fabrication approaches, and
the exciton delocalization mechanism was proposed \cite{Guo}. Very
recently, the coherent ultrafast charge transfer process was
comprehensively investigated, indicating the significant role of the
vibrational modes \cite{ScienceNew}.

Based upon all these experiments, a physical mechanism of the
coherent delocalization comes up. It indicates that, when the CT
state is formed the electron and hole rapidly separate due to the
coherent expansion of the wavefunctions in polymers. The mean
distance between electron and hole becomes much larger than that
from the theoretical expectation. Consequently the Coulomb
attraction in between is greatly reduced and the dissociation of CT
state becomes much more likely. This mechanism then explains the
high-efficient performance of BHJ photocells. In this context, we
are presently on the stage of the in-depth theoretical study on the
delocalization and coherent dynamics in organic photocells.

Theoretically, the coherent dynamics in organic semiconductors has
been widely studied in the past. The variational theory was firstly
used in the pioneer work of Silbey \cite{Silbey}, who studied the
coherent and incoherent components of the mobility very carefully.
The recent research interest in related subjects emerged in the
study of crystalline and semi-crystalline organic materials,
motivated by the development of organic field-effect transistors
based on the pentacene \cite{rmpCrystal}. In such devices over a
certain temperature extent, the mobility is found to decrease with
increasing temperature \cite{muDrop}, showing the signature of
coherent dynamics. In order to make sense of this issue Troisi
\textit{et al.} proposed the dynamic disorder model
\cite{TroisiDynamic}, in which the molecules are described by the
transport sites on a one-dimensional chain with the intersite
hopping modulated by the lattice vibrations. Following this line,
Ciuchi \textit{et al.} studied a similar system from the perspective
of the Kubo formula \cite{FratiniTransient, FratiniKubo}. Geng
\textit{et al.} then introduced the effect of high-frequency
intramolecular phonon modes into the model \cite{ShuaiDynamic}. We
have also proposed a physical picture of the decoherence effect into
the dynamic disorder model, by incorporating a decoherence time
$t_d$ \cite{Mine1} and the quantum phonon modes \cite{Mine2}. Our
very recent work tried to incorporate the formalism of Wigner
function to investigate the coherent dynamics of exciton
dissociation \cite{Mine3}.

On the other hand, an alternative way to simulate the charge
photogeneration process borrows the language of thermodynamics,
which was firstly suggested by Shockley \textit{et al.}
\cite{Shockley}. In the heat engine model they proposed, the photon
is emitted by a high-temperature emission source (\textit{e.g.}, the
sun), and the environment in which the photocell is immersed behaves
as the low-temperature sink. The photon-to-charge conversion is then
equivalent to the process that the photocell absorbs energy from the
emission source and does useful work following the loss of heat to
the environment. The recombination processes of both exciton and CT
state play essential roles in the loss of excitation energy. In
order to take the quantum coherence and delocalization into
consideration, Dorfman \textit{et al.} employed a quantum heat
engine model in which the conversion process is described
quantum-mechanically \cite{QHE}. Taking two molecules into account,
the model was adopted to compare the conversion processes in both
light-harvesting systems and semiconductors. Another work applied
the similar model with delocalized excited states to get more
efficient photogeneration \cite{Chin}, and an efficiency of about
$40\%$ is subsequently achieved. Furthermore, Wang et al. recently
utilized the quantum master equations to simulate both the electron
and heat current \cite{Cao}. The model they considered is that based
upon the double quantum dots embedding in two leads with the dot
representing the donor or acceptor molecule in photocells. The
essential physics they concerned is the inter-dot tunneling which
dominates the photovoltaic effect to a large degree. Based on their
simulations, a significant enhancement of the photovoltaic current
is obtained and the optimal value of the tunneling is determined.

In this work, we intend to follow the line of quantum heat engine
model to study the ultrafast coherent dynamics and the
delocalization in the charge photogeneration within the theoretical
framework of spin-boson model (SBM). Similar with the consideration
of \cite{Cao}, two molecules (donor and acceptor) are simulated by
two spin-halves (two-level systems). The bosonic baths coupled to
the two spins represent the emission source and the environment,
respectively. The heat to useful work conversion process in the SBM
has been extensively studied in the community of thermodynamics
\cite{SBM_heat}. But here we will employ a newly developed method
based upon the time-dependent density matrix renormalization group
(t-DMRG) algorithm \cite{mine5} and the unitary equilibration
\cite{mine6}. This method allows us to study the \textit{transient}
and \textit{coherent} energy flow through spin systems with merely
model-free assumptions. We will investigate the thermodynamics of
the model and the photogeneration process wherein. The results will
show that a steady energy current, via which we can calculate the
conversion efficiency, is quickly established following a stage of
ultrafast transient energy flow. In the extreme conditions a hyper
conversion efficiency is achieved when the quantum coherence is
quenched at the appropriate time. The argument of entanglement will
help us to understand the intrinsic physics behind. The paper is
organized as follows. The methodology we use is introduced in Sec.
II. Calculation results are presented in Sec. III, where the
transient energy flow, the conversion efficiency and the evolution
of entanglement are discussed. Conclusions are drawn in the final
section.

\section{Model and methodology}

The model we study is motivated by the quantum heat engine model
addressed by Wang \textit{et al.} recently \cite{Cao}. A quantum dot in
the model is represented by a spin-half, namely a two electronic
level system denoting the ground and excited states of the donor or
acceptor molecule, and the inter-dot tunneling between the spins
stands for the intermolecular transition at the donor-acceptor interface.
The excitons are assumed to emerge in the bulk of the donor and quickly move to the interface to form the CT states,
so two bosonic bath consisting of numerous excitons or phonons are coupling to the two spins. 

To this end, we employ the SBM \cite{review1,review2}, with two
spins sandwiched in two bosonic baths. The model Hamiltonian
reads ($\hbar\equiv1$),
\begin{eqnarray}
H&=&J\boldsymbol{\sigma}_1\cdot\boldsymbol{\sigma}_2+\sum_{
\nu=1,2}\left[\frac{\Delta}{2}\sigma^x_{\nu}+\sigma^z_{\nu}\sum_i\lambda_{i,\nu}(b^{\dag}_{i,\nu}+b_{i,\nu})\right.\nonumber\\
&+&\left.\sum_i\omega_{i,\nu}
b^{\dag}_{i,\nu}b_{i,\nu}\right],\label{hami}
\end{eqnarray}
where $\boldsymbol{\sigma}_{\nu}$ is the usual notation for Pauli
operator of the $\nu$-th spin with the superscript $x(z)$ denoting
its respective component on $x(z)$ orientation, $J$ the transition
constant, $\Delta$ the tunneling between the spin up ($\mid\uparrow\rangle$) and down ($\mid\downarrow\rangle$) along the z orientation;
$\omega_{i,\nu}$ is the frequency of the boson of $i-$th mode coupling
with the $\nu$-th spin, $\lambda_{i,\nu}$ the respective coupling
constant, and $b^{\dag}_{i,\nu} (b_{i,\nu})$ the creation
(annihilation) operator of bosons. The frequency of the boson is cut
off at $\omega_c$, such that the spectral density function is
expressed as $J_{\nu}(\omega)=2\pi\alpha\omega^{1-s}_c\omega^{s}$
with $\alpha$ being the dimensionless spin-boson coupling. For simplicity, we set $\Delta$, $\alpha$ and $s$ to be
$\nu$-independent throughout this work.

In consequence, three terms in the Hamiltonian (\ref{hami}) are
mainly concerned. The first term of the right hand side is the
transition of the energy between the two molecules, which
simulates the charge transfer process from exciton to CT state or
vice versa. The constant $J$, a similar parameter with the inter-dot
tunneling \cite{Cao}, determines the transition rate: A large $J$
gives rise to the delocalization of the CT state while the small $J$
refers to the localization. The second term denotes the energy gap
$\Delta$ between the ground state and excited state along x
orientation. With regard to the real case, the energy gap here
typically represents the binding energy of the CT state, which is of
the order $\sim100$meV. As we do not intend to consider the effect
of electric field, the $\Delta$ is set to be the same for the two
spins. The third term denotes the coupling between the spin and the
bath. A diagonal coupling (along z direction) term is employed which
implies the molecule can only be excited by the energy of the left
bath and the energy of the molecule can only be extracted by the
right bath.

\begin{figure}
\includegraphics[angle=0,scale=0.6]{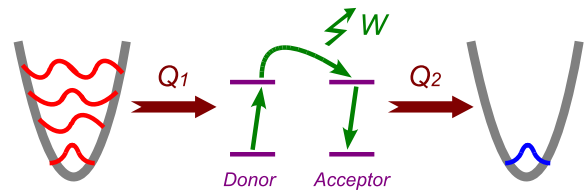}
\caption{Schematic for the photocell simulated by two spins coupling
with two baths. The two baths are simulated by bosons with different
excited energy to denote the respective temperature. The donor is excited by the high-temperature bath, then the excitation energy
may transit to the acceptor to form a CT state. The CT state could
either lose its energy to the low-temperature bath or story up the
energy for useful work. The useful work $W$ done by the photocell
then equals to the difference of the inflow and outflow heat, i.e.
$W=Q_1-Q_2$.}\label{sch}
\end{figure}

As stated, in the present model the two spins represent the donor and acceptor molecules with the state
$\mid\rightarrow\rangle(\equiv(\mid\uparrow\rangle-\mid\downarrow\rangle)/\sqrt{2})$
being the ground state of the electron and the state
$\mid\leftarrow\rangle(\equiv(\mid\uparrow\rangle+\mid\downarrow\rangle)/\sqrt{2})$
being the excited state \cite{Off}. To be specific, as shown in Fig.~\ref{sch}, we denote the
first (left) spin as the donor molecule and the second (right) one as the acceptor.
They couple to their own bath, respectively. While the device being
excited by the photon emission source, the excitons in the bulk of the donor are
generated and thus the high-temperature bath is produced. Excitons
quickly move to the interface (the two spins in our model) to form
CT states and then the useful work is made. Subsequently, an energy
current through the spins from the high-temperature bath to the
low-temperature one is established. The low-temperature bath accordingly represents the environment taking the recombination of the CT
states into account which produces the phonons in the
molecules. On the other hand, however, we
do not intend to take the respective electron current into consideration. Since the electron current is established much faster
than the heat current, it is a reasonable assumption that the
electron current does not show any qualitative influence on the heat
current \cite{Cao}.

Fig.~\ref{sch} shows the schematic of the model we study, where the
left bosonic bath is at high temperature and the right bath at low
temperature. There is an energy current flowing through the two
spins with $Q_1$ being the inflow energy from the left bath and
$Q_2$ being the outflow energy to the right bath. The difference
between inflow and outflow energy is that gained by the two spins
which will be storied up for the useful work. Thus the useful work
done by the photocell could be calculated by $W=Q_1-Q_2$, and the
conversion efficiency is
\begin{eqnarray}
\eta=\frac{W}{Q_1}=1-\frac{Q_2}{Q_1}.\label{eta}
\end{eqnarray}

In order to calculate the efficiency it is necessary to properly
calculate the inflow and outflow energy. The approaches based upon
the master equations are usually used which are convenient to deal
with the mixed-state dynamics \cite{Cao}. Alternatively, in this
work, we utilize the pure-state approach on the basis of the t-DMRG
and the orthogonal polynomials algorithm \cite{mine5}, via which the
time evolution of the SBM with diagonal spin-boson coupling could be
investigated with high precision \cite{Off}. Based on this method,
both the ultrafast charge transfer process and the efficiency with
steady state could be computed. Recently, we have studied the
unitary equilibration with the algorithm \cite{mine6}, indicating
that t-DMRG is very powerful to study the thermodynamics of the
relevant models without any model-sensitive assumptions. In
particular, with this method it is possible to investigate the
coherent and incoherent dynamics in a unified framework, rather than
to study the different components separately with the variational
theory \cite{Silbey}. Based on these preceding works, we are then
able to credibly study the thermodynamics of the quantum heat engine model and
the flowing energy wherein.

The calculating procedure is as follows. Initially we calculate the
ground state $|g\rangle$ of the Hamiltonian (\ref{hami}) by the
static DMRG method. The light-matter interaction term is then
introduced with the form as \cite{SunKW}
\begin{eqnarray}
H_{\rm l-m}=-{\rm {\bf
\hat{\mu}}\cdot\textbf{E}}=-\sum_i\vec{\mu}_i(b^{\dag}_{i,\nu}+b_{i,\nu})\cdot{\rm\textbf{E}},
\end{eqnarray}
where ${\bf \hat{\mu}}$ stands for the dipole operator with
$\vec{\mu}_i$ being the transition dipole moment for $i$-th mode,
and ${\rm\textbf{E}}$ denotes the classical radiation field. The
energy of the left bath is substantially enhanced by applying the
extern action of $e^{-iH_{\rm l-m}t_E}$ onto the ground state
$|g\rangle$ with $t_E$ being the action time of the pulsed field
which will be set to $0.5\pi$. Obviously,
$\vec{\mu}_i\cdot{\rm\textbf{E}}$ are the parameters controlling the
energy obtained from the photon emission. Here, in order to
guarantee the bath to be stable such that the subsequent results are
not sensitive to the microscopic details of the bath, the parameters
are set to be $\omega_c$ for all the modes throughout this work, so
that the energy gained by the left bath is at least two order larger
than the energy gap of the spins. Afterward, the time evolution of
the whole system is calculated by t-DMRG. In the early stage of the
evolution the flowing current of the energy will be time-dependent
and after an ultrafast process it becomes steady. Accordingly, we
target during the time evolution the energy change $\delta E_L$,
$\delta E_S$ and $\delta E_R$ for the left bath, two spins and the
right bath, respectively. Once the steady state is achieved, the
energy of the two spins will be approximately unchanged
and a steady energy flow is obtained. In this situation, $\delta
E_S/\delta E_L=(\delta E_L-\delta E_R)/\delta E_L$ goes constant
which is explicitly equal to the efficiency $\eta$, the quantity we
want to calculate.

\section{Results and discussions}

In this section, we present the calculating results we obtained for
$\omega_c=1.0, \Delta=0.1$. As we are concerning the delocalization
and the quantum coherence of the CT state, in the following
calculations we will mainly work in the coherent regime, namely the
case of the sub-Ohmic spectrum of bath. A number of works have
stated that in this regime the spin-bath coupling is relatively
strong and the spin dynamics prefers to be coherent and delocalized
\cite{SBM1,SBM2}. We will first show the transient energy flow and
based on the steady state we calculate the efficiency in terms of
delocalization. The dynamics of entanglement will be shown in the
last subsection following with the discussion of the entanglement
sudden death. We would like to emphasize here that, to some degree
our work motivated to have a proof-to-principle study to apply the
quantum heat engine model to the charge photogeneration in organic
photocells. Firstly, we do not consider the simultaneous electron
current, since we realize it does not significantly affect the heat
current due to the incongruous time scale of the two processes.
Secondly, we do not intend to explicitly connect the model
parameters we set here to that in the practical molecules. However,
the reasonable parameter extent for the organic molecules will be
kept in mind, such that the subsequent qualitative statement we
address is actually useful for the real situation.

\subsection{Transient energy flow}

\begin{figure}
\includegraphics[angle=0,scale=2.4]{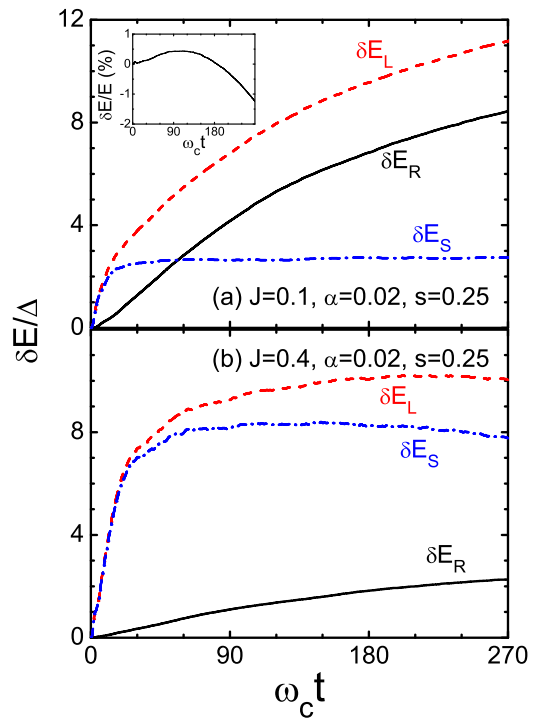}
\caption{Time evolution of energy change of the left ($\delta E_L$)
and right ($\delta E_R$) bath and the spins ($\delta E_S$). The
inset of (a) shows the relative deviation of the total energy during
the time evolution. The other parameters are: $\alpha=0.02,
s=0.25$.}\label{trans}
\end{figure}

We first show in Fig.~\ref{trans} the time evolution of the energy
change of the left and right bath and the spins for $\alpha=0.02,
s=0.25$ and two $J$'s. It is found that, the energy of the left bath
which is initially with high energy changes very quickly right after
the time evolution starts. When $\omega_ct$ is smaller than about
$50$ the $\delta E$ of the left bath and the spins are very close,
implying the spins gain a majority portion of the energy from the
emission source to form the excited state. This is the first stage
of the energy flow, and following the time advances the second stage
appears. In this stage, the right bath gains almost all the energy
emitted by the left bath, while the energy of the spins keeps nearly
unchanged during the process. The second stage is closely related to
the CT state dissociation which takes place without loss of excess
energy. The inset shows the computation error which is measured by
the relative deviation of the total energy. It is clearly that the
relative deviation is approximately smaller than $1\%$ which is in
the reasonable extent.

\begin{figure}
\includegraphics[angle=0,scale=2.4]{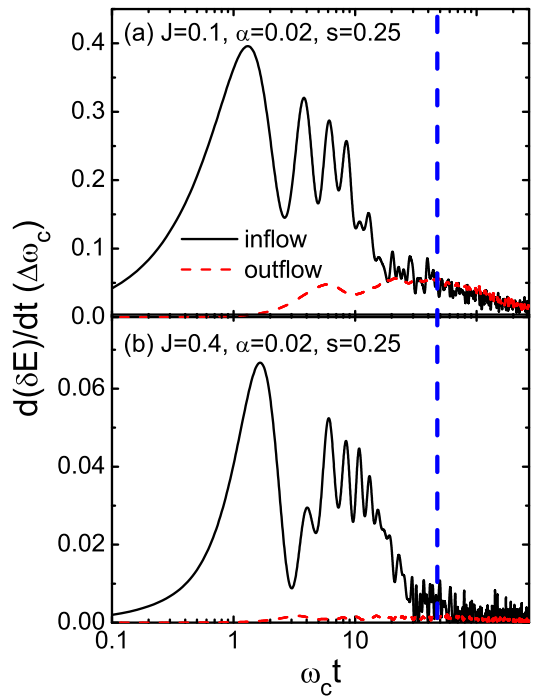}
\caption{Time evolution of the inflow and outflow energy measured by
the derivative of $\delta E$ with respect to the time. (a) and (b)
refer to $J=0.1$ and $0.4$, respectively. The blue dashed line shows
the division of the two stages of the transient energy flow. The
other parameters are: $\alpha=0.02, s=0.25$.}\label{flow}
\end{figure}

In order to show the two stages of the energy flow more clearly, we
calculate the temporal derivative of $\delta E_L$ and $\delta E_R$
which are equivalent to the energy current flowing in and out of the
spins. In Fig.~\ref{flow}, the inflow and outflow energy current are
shown for $J=0.1$ and $J=0.4$. There are clearly two stages: The
first stage is $\omega_c t<50$ during which the outflow current is
much smaller than the inflow current, and the second stage is
$\omega_c t>50$ during which the inflow and outflow current are
nearly the same. In the first stage, we find the energy current of
the left bath behaves a relatively regular oscillation, implying the
coherent resonance of the emission source and the spins. In the
second stage, however, the energy of the spins is saturated and the
lineshape of the current becomes chaotic which is a signature of the
establishment of the steady energy current.

The two-stage behavior of the transient energy flow is an essential
finding of this work. The first stage refers to the coherent
interplay between the high-temperature bath and the system, and the
second stage shows the incoherent heat flow between the two baths.
With respect to the charge photogeneration, the first stage is
mainly corresponding to the coherent photon excitation while the
second one is to the incoherent recombination. Clearly the ultrafast
delocalization/dissociation process of the CT state is closely
related to the first stage and the quantum coherence plays a
significant role wherein. This conclusion is addressed benefitting
from the advantages of our method which takes the quantum coherence
into account in a unitary manner. It is worth noting that, in our
model there is no channel for the spins to release the gained energy
and to make the useful work. The spins can only story up the energy
by changing their eigen-states and the respective populations. As we
have stated above, a reasonable assumption is that the excited
states of the spins are closely similar between the cases with and
without the electron current, especially in the presence of the
quantum coherence. So the storied energy we calculate here could be
directly related to the useful work in the real situations.

\subsection{Efficiency and delocalization}

\begin{figure}
\includegraphics[angle=0,scale=0.7]{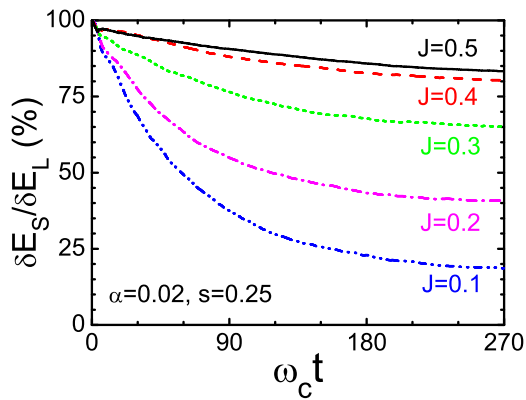}
\caption{Time evolution of the energy conversion ratio measured by
$\delta E_S/\delta E_L$ from the left bath to the spins for five
$J$'s. The other parameters are: $\alpha=0.02, s=0.25$.}\label{al02}
\end{figure}

To make a clear connection of our present model and the
photogeneration in organic molecules, we show in Fig.~\ref{al02} the
ratio of the energy conversion which measures how much energy from
the emission source is transferred to the spins. It is found that,
once the steady state is achieved, for the case $J$ is small
($J=0.1$) the spins gain a tiny portion (around $20\%$) of the
energy emitted from the left bath, while when $J$ becomes large
($J=0.5$), the spins gain the majority (around $85\%$) of the
emitted energy. This effect is easy to understand, since the
transition term of Hamiltonian (\ref{hami}) opens a large gap of
singlet and triplet states of the two spins depending on the value
of $J$. When $J$ is sufficiently large, a single excited electron
will carry more energy than that of the small $J$ case. Thus the
excited state in the large $J$ case could be referred as ``hot"
state. Subsequently, this effect is due to the delocalization of the
spins' state which simulates the real situation that the delocalized
CT state is ``hot" with more energy which prefers to dissociate
rather than the localized CT state.

\begin{figure}
\includegraphics[angle=0,scale=0.7]{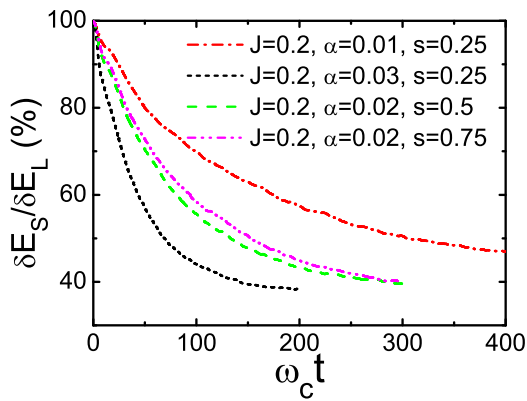}
\caption{Time evolution of the energy conversion ratio measured by
$\delta E_S/\delta E_L$ from the left bath to the spins for four
sets of parameters.}\label{al0103}
\end{figure}

\begin{figure}
\includegraphics[angle=0,scale=0.7]{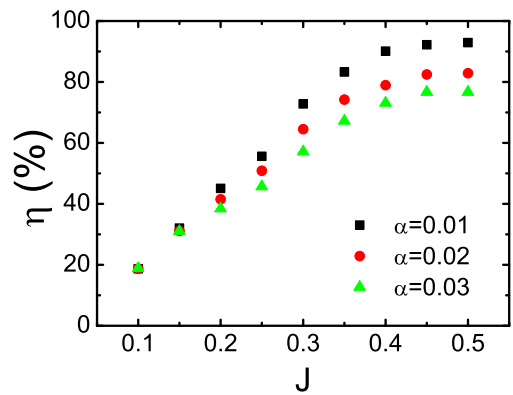}
\caption{Efficiency of the energy conversion versus $J$ for $s=0.25$
and three $\alpha$'s.}\label{Eta}
\end{figure}

In Fig.~\ref{al0103} we show the ratio for another four sets of
parameters, that is, to keep $J$ unchanged and adjust $\alpha$ and
$s$. The parameters $\alpha$ and $s$ determine the coupling between
the system and the bath and the bath's spectrum, respectively. As
stated, in an equilibrated system the physical property is expected
to be independent of the microscopic details of the bath's spectrum.
Our results prove this statement, as the two curves of the ratio
evolution for various $s$ are almost identical. On the other hand,
when we adjust the coupling strength $\alpha$, it shows that the
steady state is more likely to be obtained for $\alpha=0.03$ rather
than $\alpha=0.01$. This is easy to understand as the large coupling
induces a quick quench of the quantum coherence of the spins. More
importantly, for all the three $\alpha$'s the conversion ratios seem
to decay to the similar value, implying that the efficiency is only
weakly dependent of the coupling. The conversion efficiency shown in
Fig.~\ref{Eta} supports this result, as we can see $\eta$ changes a
little when $\alpha$ changes. There is still a small difference
which is, as we recognize, due to the limitation of the numerical
method. It is also worth noting that, we do not show the long-term
evolution for the case of large $\alpha$ because the numerical
precision decays quickly with increasing $\alpha$ \cite{mine5}.

In Fig.~\ref{Eta} the efficiency of energy conversion is shown which
is measured by the steady value of $\delta E_S/\delta E_L$. It is
clearly that the efficiency increases following increasing $J$ which
has been discussed above. The largest efficiency for $\alpha=0.01$
is about $93\%$, far beyond the value obtained by the master
equations calculations \cite{Chin}. This remarkable enhancement of
efficiency is mainly originated from the fact that the
low-temperature (right) bath is initially at the ground state, i.e.
zero temperature. According to the formula of the Carnot efficiency,
it should be this large. Meanwhile, the quench of the coherence
takes places at the appropriate time benefitting the efficiency
enhancement, as from the energy flow shown in Fig.~\ref{flow} we can
find that once the coherence is quenched the energy flow will be
greatly reduced. In this situation the energy storied in the spins
is not easy to flow out. While in the treatment of the master
equations, there is always an available incoherent channel for the
outflow energy current to suppress the conversion efficiency. This
explains the enhancement of efficiency in the present work compared
with the previous one with the quantum heat engine model. In
addition, we find that the efficiency undergoes a saturation with
increasing $J$. This is because the energy gap induced by $J$ is
much larger than the cutoff frequency of the bosons.

As to the real situation, it is evident in the light-harvesting
systems that the efficiency reaches to $100\%$, as we have addressed
above. Hence, our result seems more closely to the light-harvesting
case. On the contrary, however, in the organic photocells an
incoherent loss of excitation energy is quite possible with respect
to the nongeminate recombination of charge carriers. So the
efficiency in these systems is much smaller than that of the
theoretical expectation. In order to improve the efficiency wherein,
it is thus necessary to optimize the molecular and device structure
to appropriately handle the quantum coherence. As suggested in
\cite{Cao}, with sufficient optimization the theoretical limit of
efficiency we obtain here is reachable.

\subsection{Role of entanglement}

The entanglement of the two spins is a useful measurement of the
quantum coherence and delocalization. In the presence of the
system-bath coupling, the entanglement of the spins would be easy to
be quenched by the two baths. It is apparent that the longer the
entanglement survives, the stronger the delocalization effect is.
Furthermore, we expect that the survival of the entanglement should
be consistent with the transition of the two stages of the transient
energy flow.

\begin{figure}
\includegraphics[angle=0,scale=2]{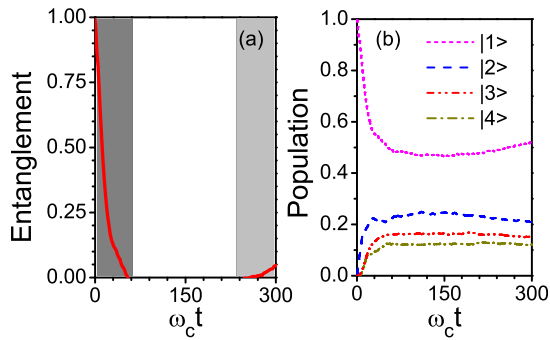}
\caption{(a) Time evolution of the entanglement between the two
spins. The dark and light areas show the regimes of the entanglement
sudden death and revival. (b) Time evolution of the population for
the four eigenstates of the two-spins' reduced density matrix. The
state $\mid n=1,2,3,4\rangle$ is labeled by the descending order of
the population. The parameters are: $J=0.4, \alpha=0.02,
s=0.25.$}\label{Ent}
\end{figure}

In this work, the entanglement is measured by the concurrence
\cite{Conc}. Fig.~\ref{Ent}(a) shows the time dependence of the
quantum entanglement between the two spins for $J=0.4, \alpha=0.02,
s=0.25$. It is shown that there appears an entanglement sudden death
(ESD) effect at around $\omega_c t=50$. The effect is synchronous
with the transition of the two stages of energy flow, so that it
proves our statement above. This two-stage transition can also be
seen in the population evolution of the four eigen-states of the
spins' reduced density matrix, which is shown in Fig.~\ref{Ent}(b).
Initially, the spins are residing in their ground state, and
following the time advances the population will be quickly transited
to the other three states. This transition occurs very quickly and
after that the populations keep nearly unchanged during the
long-term evolution. In addition, there is a revival of the
entanglement after about $\omega_c t=240$. This is also observed in
the zero-temperature dynamics \cite{mine5}, which should be related
to the robust non-Markovian feature of the bath. Consequently, we
would like to address the statement that, the enhancement of the
conversion efficiency is strongly dependent of the quench of the
non-Markovian entanglement in the process of photogeneration.

\begin{figure}
\includegraphics[angle=0,scale=0.7]{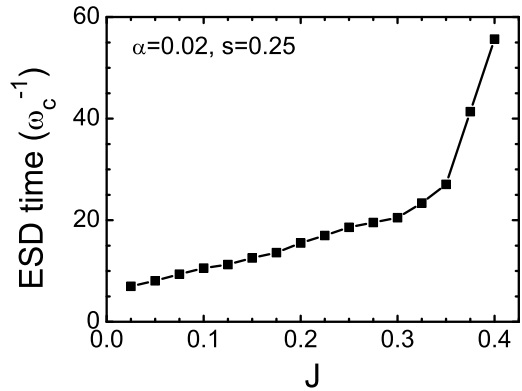}
\caption{The time point of entanglement sudden death versus $J$ for
$\alpha=0.02, s=0.25$.}\label{ESD}
\end{figure}

As discussed above, the ESD effect we observe here is essential for
the charge photogeneration process, since it is a criterion for
dividing the coherent (ultrafast) and incoherent dynamics. Hence, it
is useful for us to plot the time point of the ESD in terms of $J$,
which is shown in Fig.~\ref{ESD}. We can find that when $J<0.35$ for
$\alpha=0.02$, the time of ESD increases linearly with increasing
$J$. For larger $J$, the time of ESD diverges very quickly, and when
$J>0.4$ the entanglement can not be quenched at all. The point of
division here is very similar with the point for the saturation of
the $\eta$ as shown in Fig.~\ref{Eta}. This means the survival of
the quantum coherence plays an essential role in the conversion
efficiency. Thus, the ESD time we compute here is a useful criterion
for the estimate of delocalization and photogeneration rate in the
real molecules.

\section{Conclusion}

In summary, we have made use of the sub-Ohmic spin-boson model to
simulate the charge photogeneration process in organic molecules.
The two spins denote the donor and acceptor molecules and the two
baths represent the emission source and the environment,
respectively. The transition constant $J$ between the two spins
refers to the degree of delocalization which is very important for
the photogeneration. In our calculations, a two-stage transition of
the energy flow is obtained. The first stage refers to an ultrafast
energy transition process from the emission source to the spins.
Following that, in the second stage a steady energy flow from source
to environment is achieved in a completely unitary manner. This
effect should be contributive to the research field of the heat
current through a spin system. On the other hand, our result shows
that the conversion efficiency is strongly dependent of $J$, the
delocalization, and insensitive to the system-bath coupling and the
bath's spectrum. The evolution of the entanglement between the two
spins shows its sudden death at short time duration and long-termly
revival. The sudden death of the entanglement is synchronous with
the transition of the two stages of the transient energy flow,
implying the essential role of the quantum coherence in the
transition. Finally, the time point of the sudden death is
recognized to be a helpful quantity for the study of the
photogeneration.

Two important concluding remarks are worth making. Firstly, the
transition constant $J$ denoting the effect of delocalization is the
most important parameter in our present model. If the binding energy
of the CT state is $100$meV related to $\Delta=0.1$, the chosen
value of $J$ in this work is equivalent to about several hundred
meV. This quantity is available in the organic small molecules with
good crystallinity or the polymers with long conjugated length. In
these materials the delocalization can not be neglected, and our
results give the limit of the energy conversion efficiency taking
the quantum coherence into account. Secondly, our results show the
bath's spectrum does not affect the photogeneration efficiency. This
is theoretically predictable and suggests that in experiment the
specific absorption spectrum of the materials does not matter in the
charge photogeneration.

\begin{acknowledgments}
This work was supported by the NSF of China (Grant Nos. 91333202,
11134002 and 11104035) and the National Basic Research Program of
China (Grant No. 2012CB921401).
\end{acknowledgments}


\begin{thebibliography}{99}
\bibitem{Off} Via an approapriate rotation operation, the Hamiltonian (\ref{hami}) could be transformed to that with off-diagonal coupling term.
As we have tested, however, the t-DMRG algorithm is much more
accurate to deal with the diagonal coupling rather than the
off-diagonal coupling. Without loss of generality, in the present
work the states along x orientation are regarded to be responsible
for the ground state and excited state in the real molecules.
\end{thebibliography}
\end{document}